\documentclass[aps, pra, twocolumn, superscriptaddress, amsmath,  tightenlines, longbibliography]{revtex4-1}

\usepackage{dcolumn}
\usepackage{graphicx,CJK}
\usepackage{epstopdf}
\usepackage{mathrsfs}
\usepackage{subfigure}
\usepackage{booktabs}
\usepackage{amsmath}
\usepackage{physics}
\usepackage{dsfont}
\usepackage{amstext}
\usepackage{amssymb}
\usepackage{amsbsy}
\usepackage{bbm}
\usepackage{amsthm}
\usepackage{graphicx}
\usepackage{xcolor}
\usepackage[colorlinks,urlcolor=blue,linkcolor=blue,citecolor=blue]{hyperref}

\usepackage{url}
\usepackage[colorlinks]{hyperref}
\hypersetup{%
	plainpages=true,
	breaklinks=true,       
	hypertexnames=false,  
	pageanchor=true,
	colorlinks=true,
	linkcolor={blue},
	citecolor={blue},
	urlcolor={blue},
	anchorcolor={black}
}

 \makeatletter

\newcommand{\Rmnum}[1]{\expandafter\@slowromancap\romannumeral #1@}
\makeatother

\begin{document}
\title{Anomalously Reduced Homogeneous Broadening of Two-Dimensional Electronic Spectroscopy at High Temperature by Detailed Balance}
\author{Ru-Qiong Deng}
\affiliation{Department of Physics, Applied Optics Beijing Area Major Laboratory, Beijing Normal University, Beijing 100875, China}
\affiliation{Key Laboratory of Multiscale Spin Physics, Ministry of Education, Beijing Normal University, Beijing 100875, China}
\author{Cheng-Ge Liu}
\affiliation{Department of Physics, Applied Optics Beijing Area Major Laboratory, Beijing Normal University, Beijing 100875, China}
\affiliation{Key Laboratory of Multiscale Spin Physics, Ministry of Education, Beijing Normal University, Beijing 100875, China}
\author{Yi-Xuan Yao}
\affiliation{Department of Physics, Applied Optics Beijing Area Major Laboratory, Beijing Normal University, Beijing 100875, China}
\affiliation{Key Laboratory of Multiscale Spin Physics, Ministry of Education, Beijing Normal University, Beijing 100875, China}
\author{Jing-Yi-Ran Jin}
\affiliation{Department of Physics, Applied Optics Beijing Area Major Laboratory, Beijing Normal University, Beijing 100875, China}
\affiliation{Key Laboratory of Multiscale Spin Physics, Ministry of Education, Beijing Normal University, Beijing 100875, China}
\author{Hao-Yue Zhang}
\affiliation{Department of Physics, Applied Optics Beijing Area Major Laboratory, Beijing Normal University, Beijing 100875, China}
\affiliation{Key Laboratory of Multiscale Spin Physics, Ministry of Education, Beijing Normal University, Beijing 100875, China}
\author{Yin Song}
\email{songyin2021@bit.edu.cn}
\affiliation{School of Optics and Photonics, Beijing Institute of Technology, Beijing 100081, China}
\author{Qing Ai}
\email{aiqing@bnu.edu.cn}
\affiliation{Department of Physics, Applied Optics Beijing Area Major Laboratory, Beijing Normal University, Beijing 100875, China}
\affiliation{Key Laboratory of Multiscale Spin Physics, Ministry of Education, Beijing Normal University, Beijing 100875, China}
\date{\today}

\begin{abstract}
Dissipation and decoherence of quantum systems in thermal environments are important to various spectroscopies. It is generally believed that dissipation can broaden the line shape of spectroscopies and thus stronger system-bath interaction can result in more significant homogeneous broadening of two-dimensional electronic spectroscopy (2DES). Here we show that the case can be the opposite in the regime of electromagnetically induced transparency (EIT). We predict that assisted by the EIT, the homogeneous broadening of the 2DES at a higher temperature can be significantly reduced due to the detailed balance.
This anomalous effect is due to the long-lasting off-diagonal peaks in 2DES.
\end{abstract}
\maketitle

\section{\label{sec:Intro}Introduction}
As known to all, multitransition and higher temperature can induce more significant decoherence and thus result in broader line width \cite{Mukamel1999,Breuer2002}. However, it has been theoretically predicted and experimentally observed that the multitransition of a nitrogen-vacancy (NV) center in diamond can have longer coherence time than the single transitions, due to manipulation of the quantum bath evolution via flips of the center spin \cite{Zhao2011PRL,Huang2011NC}. Therefore, it might be interesting to investigate the effect of the temperature on the homogeneous broadening in spectroscopies. In this paper, we theoretically demonstrate that the homogeneous broadening of the two-dimensional electronic spectroscopy (2DES) in the presence of the electromagnetically-induced transparency (EIT) can be anomalously reduced at higher temperatures because of the detailed balance.

{\color{red}
In recent years, two-dimensional electronic spectroscopy (2DES) has emerged as a powerful tool for investigating the ultrafast dynamics of complex quantum systems \cite{Mukamel1999,Weng2013,Cho08CR,Mukamel00ARPC,Chernyak95PRL,Jonas03ARPC,Biswas22CR,Hamm2011,Moody17APX,Fuller15ARPC}, including quantum wells \cite{Turner10Nature}, quantum dots and 2D materials \cite{Liu19PRL,Moody15NC}, perovskites \cite{Liu21SA}, organic photovoltaic cells \cite{Song14NC,Bakulin16NC,Jones20NC,Sio16NC}, photosynthetic complexes \citep{Brixner2005Nature,Engel07Nature,Collini10Nature,Romero14NP,Fuller14NC,Tiwari13PNAS,Cao20SA,Song21NC,Son19C}, NV centers in diamond \citep{Huxter2013NP}, and chiral molecules \cite{Cao2022PRL}. The fundamental theoretical framework of 2DES involves the application of three coherent laser pulse trains to samples, generating third-order polarization signals. Through the analysis of the nonlinear signals, 2DES can unveil nuanced aspects of electronic coherence, energy transfer pathways, and correlations within a wide array of physical systems. Notably, explorations of ultrafast processes in molecular aggregates, photosynthetic complexes, and semiconductor nanostructures has unearthed fundamental principles governing the dynamics of excited states. Given that the time-correlation function intricately influences quantum dynamics in conjunction with the system Hamiltonian, the center-line slope of 2DES has been proposed as a means to extract information regarding system-bath interaction under various conditions \citep{Kwak2007JCP,Rosenfeld2011S,Sun2023AQT}. When spectral line bands from different processes overlap, making it challenging to distinguish distinct peaks \cite{Roberts06JCP}, the ability of 2DES to resolve both the structure and dynamics is significantly hindered. The problem tends to get worse as the environment around a molecule becomes more diverse. As a result, the application of 2DES is significantly limited by the spectral line width.}

On the other hand, the EIT has been widely used to realize optical non-reciprocity \cite{Zhang2018NP,Huang2022AdP}, suppress dissipation in artificial light-harvesting systems \cite{Dong2012LSA}, and polarize the nuclear spin at the vicinity of NV centers in diamond \cite{Dutt07Science,Wang2018PRA}. Intuitively, it may be natural to utilize the EIT to effectively improve the signal-noise ratio of 2DES \citep{Liu2020JPCL}. The previous investigation proposes utilizing the quantum optical EIT effect to improve multi-dimensional spectroscopic measurements beyond the standard resolution limits \cite{Liu2020JPCL}. However, at high temperature, only the downhill population relaxation has been taken into account, while the uphill population relaxation has been neglected therein. In other words, the whole relaxation does not fulfill the detailed-balance condition \cite{Mukamel1999,Breuer2002}. But it is worth noting that under non-zero temperature conditions, the uphill population relaxation in combination with the downhill population relaxation drive the open quantum system towards the steady state described by the Boltzmann distribution \cite{Mukamel1999,Breuer2002}. It is this fundamental principal that inspires us to take into account this overlooked crucial information and consider the influence of temperature on the spectral resolution, making it more aligned with real-world scenarios.

{\color{red}The rest of the paper is structured as follows. In Sec.~\ref{sec:Model}, we introduce the theoretical model and perform calculations based on the response functions. In Sec.~\ref{sec:Results}, numerical simulation results are presented and analyzed. Sec.~\ref{sec:Implementation} describes the physical implementation. Finally, a summary and discussion are provided in Sec.~\ref{sec:Conclusion}. In Appendixes~\ref{sec:appendixA} and \ref{sec:appendixB}, we derive the master equations for a two-level system and a four-level system, respectively. In Appendix~\ref{sec:appendixC}, we discuss the whole rephasing signal.}

\begin{figure}
\includegraphics[width=8cm]{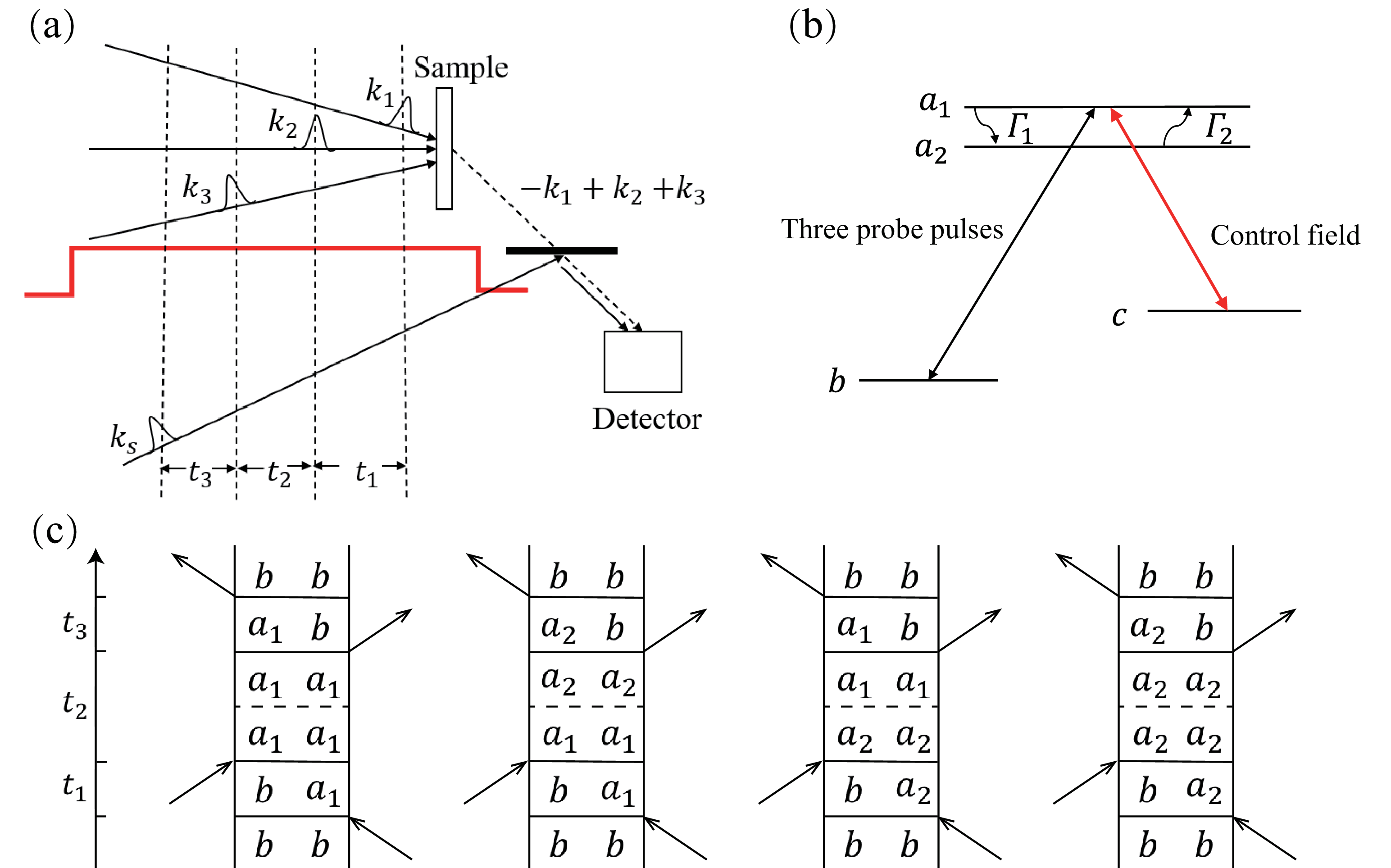}
\caption{Schematic illustration of removing homogeneous broadening by EIT and coupling to a high-temperature bath: (a) Pulse sequence, (b) four-level system, and (c) Feynman diagrams.}
\label{fig:scheme}
\end{figure}

\section{Theoretical Model}
\label{sec:Model}

In essence, the two-dimensional spectroscopy is the
third-order nonlinear polarization signal generated by the interaction
of three coherent ultra-short laser pulses with the substance \cite{Mukamel1999}, where the heterodyne detection method is applied to detect the signal. As
shown in Fig.~\ref{fig:scheme}(b), four pulses are applied, including three short probe pulses and one long control field. The first probe pulse
is temporally centered at $t=0$, and the next two probe pulses are delayed by $t_{1}$ and $t_{2}$ in time successively. Finally, a heterodyne pulse is applied after the time interval $t_{3}$. The first three ultra-short laser pulses generate a signal, which is heterodyne detected by a fourth pulse in a specific phase-matching direction. In 2DES, excitation frequency
axis $\omega_1$ and detection frequency axis $\omega_3$ are the Fourier transform of the time delays $t_{1}$ and $t_{3}$, respectively. In addition to a sequence of four pulses
commonly-used in traditional two-dimensional spectroscopy, a
narrow-band control field is applied to drive a specific transition between the energy levels $a_{1}$ and $c$ in Fig.~\ref{fig:scheme}(b). Here, broadband probe pulses can drive
the transitions between $b$ and $a_{j}$ $(j=1,2)$.

Without loss of generality, we first consider the 2DES generated by the $R_{2}$ path, that is, the stimulated emission process in the direction of $k_{s}=-k_{1}+k_{2}+k_{3}$. Here, four Feynman diagrams can be drawn for the $R_{2}$ path,
as shown in Fig.~\ref{fig:scheme}(c). We remark that in the three-level case, i.e., $a_{2}$ is absent, there is only one kind of Feynman diagram for the $R_{2}$ path, which is not shown here. And the broadband probe pulses can not induce the transitions $a_{1}\rightleftharpoons a_{2}$ and $c\rightleftharpoons a_{j}$, but $b\rightleftharpoons a_{j}$ $(j=1,2)$.

The total Hamiltonian including the interaction between the system and the control field reads
\begin{eqnarray}
H & = & \sum_{j}\omega_{j}|j\rangle\langle j|-\frac{\Omega}{2} e^{-i\nu_{c}t}|a_{1}\rangle\langle c|+\mathrm{h.c.},
\end{eqnarray}
where $\omega_{j}$ is the energy of the state $|j\rangle$ $(j=b,a_{1},a_{2},c)$,
we assume $\hbar=1$ for simplicity, $\Omega=\mu_{a_{1}c}\varepsilon_{c}$
is the Rabi frequency with $\mu_{a_{1}c}$ being the transition dipole
moment, $\varepsilon_{c}$ and $\nu_{c}$ being the amplitude and
the frequency of the control field, respectively. Here, $\omega_{a_{2}}$ is adjustable in our investigation.

Generally, the quantum dynamics are governed by the quantum master equation $\dot{\rho}=-\frac{i}{\hbar}[H,\rho]-\varGamma_1\mathcal{L}(A_{a_2a_1})\rho-\varGamma_2\mathcal{L}(A_{a_1a_2})\rho$
\citep{Olaya-Castro2008PRB,Mukamel1999,Dong2012LSA}, where $\mathcal{L}(A_{\alpha\beta})\rho=\frac{1}{2}\{A^\dagger_{\alpha\beta}A_{\alpha\beta},\rho\}- A_{\alpha\beta}\rho A^\dagger_{\alpha\beta}$ with $\{A^\dagger_{\alpha\beta}A_{\alpha\beta},\rho\}$ being the anti-commutator, $A_{\alpha\beta}=|\alpha\rangle\langle\beta|$ is the quantum jump operator from the initial state $|\beta\rangle$ to the final state $|\alpha\rangle$. The exact quantum dynamics can be obtained by the hierarchical equation of motion, which can be exponentially accelerated by a recently-developed quantum algorithm \cite{Wang2018NPJQI,Zhang2021FoP}. However, under certain circumstances, the quantum master equation approach without the quantum-jump term can provide an analytical result and thus effectively help us grasp the underlying mechanism. In the interaction picture, assuming $\nu_{c}=\omega_{a_{1}}-\omega_{c}$, the time evolution of the density matrix is determined by
\begin{equation}
\begin{split}
\dot{\tilde{\rho}}_{a_{1}a_{1}} & =  -\varGamma_{1}\tilde{\rho}_{a_{1}a_{1}}+\varGamma_{2}\tilde{\rho}_{a_{2}a_{2}}+\frac{i}{2}\Omega\left(\tilde{\rho}_{ca_{1}}-\tilde{\rho}_{a_{1}c}\right),\\
\dot{\tilde{\rho}}_{a_{1}c} & =  \frac{i}{2}\Omega\left(\tilde{\rho}_{cc}-\tilde{\rho}_{a_{1}a_{1}}\right)-\gamma_{a_{1}c}\tilde{\rho}_{a_{1}c},\\
\dot{\tilde{\rho}}_{cc} & =  \frac{i}{2}\Omega\left(\tilde{\rho}_{a_{1}c}-\tilde{\rho}_{ca_{1}}\right)-\varGamma_{c}\tilde{\rho}_{cc}, \\
\dot{\tilde{\rho}}_{a_{2}a_{2}} & =  \varGamma_{1}\tilde{\rho}_{a_{1}a_{1}}-\varGamma_{2}\tilde{\rho}_{a_{2}a_{2}},
\end{split}
\end{equation}
where $\gamma_{a_{1}c}=(\varGamma_{1}+\varGamma_{c})/2+\gamma^{(0)}_{a_{1}c}$, the population relaxation rates between the states $a_{1}$ and $a_{2}$ are respectively $\varGamma_{1}$ and $\varGamma_{2}$, between which the relation is governed by the detailed balance \citep{Breuer2002}. $\gamma^{(0)}_{a_{1}c}$ is the pure-dephasing rate between states $a_1$ and $c$. Here, we have assumed that the population relaxation $\varGamma_{c}$ of the metastable state $c$ can be neglected. As a result, we can obtain the Green function as
\begin{eqnarray}
\mathcal{G}_{a_{1}a_{1},a_{1}a_{1}} & = & \frac{-1}{4iA_{1}A_{3}\widetilde{\Omega}}\left[A_{1}e^{-\frac{\gamma_{a_{1}c}}{2}t}\left(A_{2}^+e^{i\frac{\widetilde{\Omega}}{2}t}+A_{2}^-e^{-i\frac{\widetilde{\Omega}}{2}t}\right)\right.\nonumber \\
 &  & \left. -2i\widetilde{\Omega}\left(\varGamma_{2}A_{3}+\varGamma_{1}A_{4}e^{-A_{1}t}\right)\right],\nonumber \\
\mathcal{G}_{a_{2}a_{2},a_{1}a_{1}} & = & \frac{\varGamma_{1}}{4i\widetilde{\Omega}A_{1}B_{5}}\left[2i\widetilde{\Omega}B_{1}\left(e^{-A_{1}t}-B_{2}\right)\right.\nonumber \\
 &  &\left. +A_{1}B_{3}e^{-\frac{\gamma_{a_{1}c}}{2}t}\left(e^{i\frac{\widetilde{\Omega}}{2}t}+B_{4}e^{-i\frac{\widetilde{\Omega}}{2}t}\right)\right],\\
\mathcal{G}_{a_{1}a_{1},a_{2}a_{2}} & = & -\frac{\varGamma_{2}}{A_{1}}\left(-1+e^{-A_{1}t}\right),\nonumber \\
\mathcal{G}_{a_{2}a_{2},a_{2}a_{2}} & = & \frac{1}{A_{1}}\left(\varGamma_{1}+\varGamma_{2}e^{-A_{1}t}\right),\nonumber
\end{eqnarray}
where $\widetilde{\Omega}=\sqrt{4\Omega^{2}-\gamma_{a_{1}c}^{2}}$, $A_{1}=\varGamma_{1}+\varGamma_{2}$,
$A_{2}^\pm=[\varGamma_{2}(\varGamma_{1}+\varGamma_{2}-\gamma_{a_{1}c})+\Omega^2](i\widetilde{\Omega}\pm\gamma_{a_{1}c})\mp2\varGamma_{1}\Omega^2$,
$A_{3}=-(\varGamma_{1}+\varGamma_{2})(\varGamma_{1}+\varGamma_{2}-\gamma_{a_{1}c})-\Omega^{2}$, $A_{4}=2A_{3}+\Omega^{2}$, $B_1=-2 (\varGamma_1+\varGamma_2 ) (\varGamma_1+\varGamma_2-\gamma_{a_1 c} )-\Omega^2$, $B_2=( B_1-\Omega^2 )/2 $, $ B_3=( \varGamma_1+\varGamma_2-\gamma_{a_1 c} ) ( \gamma_{a_1 c}+i \widetilde{\Omega} )+2\Omega^2 $ and $ B_4=( \varGamma_1+\varGamma_2-\gamma_{a_1 c} ) ( -\gamma_{a_1 c}+i \widetilde{\Omega} )-2\Omega^2 $.

If we consider the $R_{2}$ term in the rephrasing case, the response function is written as \cite{Mukamel1999}
\begin{eqnarray}
S(\omega_{3},t_{2},\omega_{1}) & = & \mathsf{\textrm{Re}}\mathop{\sum_{i,j=1}^{2}|\mu_{ba_{i}}|^{2}|\mu_{a_{j}b}|^{2}\mathscr{\mathcal{G}}_{a_{j}b,a_{j}b}}(\omega_{3})\nonumber \\
 &  & \times\mathscr{\mathcal{G}}_{a_{j}a_{j},a_{i}a_{i}}(t_{2})\mathscr{\mathcal{G}}_{ba_{i},ba_{i}}(\omega_{1}),
\end{eqnarray}
where $\mu_{ba_{i}}$ is the transition dipole moment between the states
$b$ and $a_{i}$ ($i=1,2$).

\begin{figure}
\includegraphics[bb=0 0 595 535,width=8.5cm]{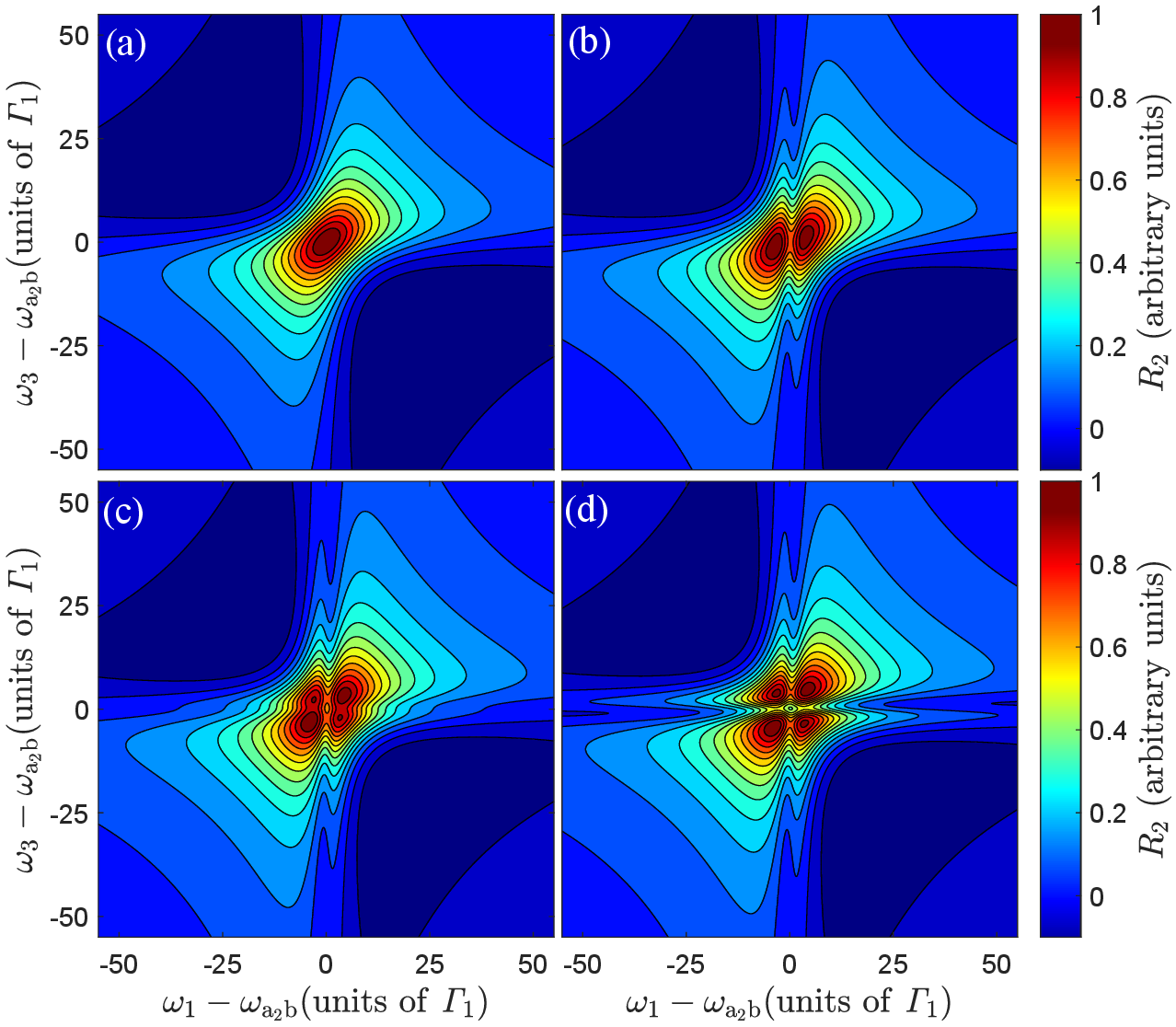}
\caption{2DES with $\omega_{a_{1}b}-\omega_{a_{2}b}=0.18\varGamma_{1}$ when $t_2=3\varGamma_1^{-1}$, and (a) $k_BT=0.01\varGamma_{1}$, $\Omega=0$,
(b) $k_BT=0.01\varGamma_{1}$, $\Omega=9\varGamma_{1}$,
(c) $k_BT=0.1\varGamma_{1}$, $\Omega=9\varGamma_{1}$,
(d) $k_BT=5\varGamma_{1}$, $\Omega=9\varGamma_{1}$.\label{fig:NearResonant}}
\end{figure}

\section{Numerical Results and Analysis}
\label{sec:Results}

We investigate the effect of temperature on the 2DES under the near-resonant condition in the long-population-time limit, as shown in Fig.~\ref{fig:NearResonant}.
When there is no control field, there is only one peak in Fig.~\ref{fig:NearResonant}(a)
due to the absence of EIT. If the control field is applied, e.g. Fig.~\ref{fig:NearResonant}(b),
the diagonal peak is split into two peaks, in which the homogeneous broadening is partially reduced. However, if the temperature is increased, e.g. $k_BT=0.1\varGamma_1$
in Fig.~\ref{fig:NearResonant}(c), two additional small peaks begin to emerge at the side of the above two peaks. Interestingly, if temperature is sufficiently high, i.e., $k_BT=5\varGamma_1$ in Fig.~\ref{fig:NearResonant}(d),
the original large peak in Fig.~\ref{fig:NearResonant}(a) is almost evenly split into four peaks. Notice that
the homogeneous broadening is even narrower than that in Fig.~\ref{fig:NearResonant}(b),
which has been significantly reduced due to the EIT. In general, the downhill and uphill rates fulfill the detailed balance \citep{Breuer2002}.
At the absolute zero temperature, there is only population transfer from the higher level to the lower one. As the temperature increases, the population-back transfer emerges due to the heating by the bath.
We remark that the combination of the heating and the EIT results
in the eliminating of homogeneous broadening in the long-population-time limit.

\begin{figure*}
\includegraphics[width=18cm]{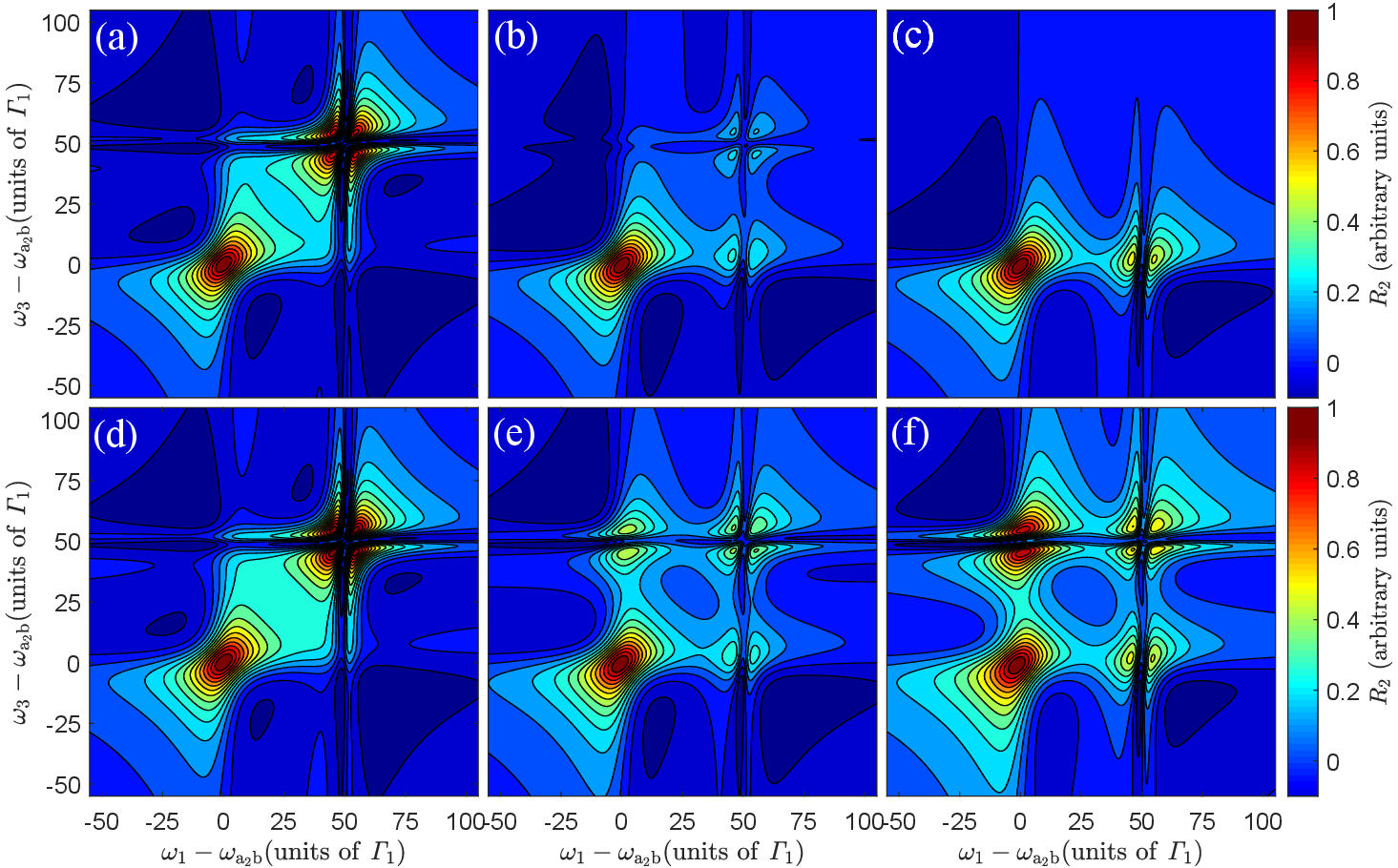}
\caption{2DES for $\Omega=9 \varGamma_{1}$,  $\omega_{a_{1}b}-\omega_{a_{2}b}=50\varGamma_{1}$ and $t_2=$ (a,d) $0$, (b,e) $0.5\varGamma_1^{-1}$, (c,f) $5\varGamma_1^{-1}$. In the top panel,
$k_BT=\varGamma_{1}$, while in the bottom
$k_BT=10^3\varGamma_{1}$. \label{fig:LargeDetuned}}
\end{figure*}

In order to illustrate the underlying physical mechanism explicitly, we consider the case of a large detuning between the
two levels $a_{1}$ and $a_{2}$. We discuss the $k_BT=\varGamma_1$
case as shown in the upper panel of Fig.~\ref{fig:LargeDetuned},
i.e., at low temperatures. When $t_2=0$, there are two sets
of diagonal peaks at $\omega_{a_{j}b}$ ($j=1,2$), respectively.
The diagonal peak at $\omega_{a_{1}b}$ is split into four small
peaks due to the EIT introduced by the control field, which induces the transition between $a_{1}$ and $c$, while the diagonal peak at $\omega_{a_{2}b}$ remains as a whole large peak. When the population time elapses, e.g. $t_2=0.5\varGamma_1^{-1}$, a set of off-diagonal peaks at ($\omega_{a_{1}b},\omega_{a_{2}b}$) emerge due to the population transfer from level $a_{1}$ to $a_{2}$. And generated by the processes corresponding to the four Feynman diagrams in Fig.~\ref{fig:scheme}(c), the peaks in 2DES which overlap with each other in the nearly-resonant case will be separated. When the population time is sufficiently long, e.g. $t_2=5\varGamma_1^{-1}$, we can observe from the spectroscopy that the peaks at $\omega_3=\omega_{a_{1}b}$ have disappeared and the peak in the bottom right corner has been enhanced. And that's because in the long-population-time limit, i.e., $t_2\gg\varGamma_{1}^{-1}$, the entire population of the level $a_{1}$ has been unidirectionally transferred
to the level $a_{2}$. The lower panel of Fig.~\ref{fig:LargeDetuned}
is simulated at $k_BT=10^3\varGamma_{1}$, that is, at a sufficiently-high temperature. In the case of $t_2=0$, we can not discriminate the difference between Fig.~\ref{fig:LargeDetuned}(d) for $k_BT=10^3\varGamma_{1}$ and Fig.~\ref{fig:LargeDetuned}(a) for $k_BT=\varGamma_{1}$. When $t_2=0.5\varGamma_1^{-1}$, two off-diagonal peaks appear in Fig.~\ref{fig:LargeDetuned}(e). In addition to the one in Fig.~\ref{fig:LargeDetuned}(b), there is one at ($\omega_{a_{2}b},\omega_{a_{1}b}$)
because of the population-back transfer from the lower level to the higher level due to heating by the bath. At $t_2=5\varGamma_1^{-1}$, the peaks in these four regions of Fig.~\ref{fig:LargeDetuned}(f) do not disappear, which is quite different from the observation in the case of $k_BT=\varGamma_{1}$. Because of non-vanishing $\varGamma_{2}$, the bidirectional population transfer between levels $a_{1}$ and $a_{2}$ always exists dynamically.

In a real system, there is not only the downhill relaxation from $a_{1}$ to $a_{2}$, but also the uphill relaxation from $a_{2}$ to $a_{1}$. The ratio of the two relaxation rates is determined by the temperature and the energy gap between $a_{1}$ and $a_{2}$. In order to observe the different scenarios as shown in Figs.~\ref{fig:NearResonant}-\ref{fig:LargeDetuned},
we can effectively adjust the energy gap between $a_{1}$ and $a_{2}$. As the energy gap increases, the 2DES
obtained when $a_{1}$ and $a_{2}$ are nearly-resonant will split,
and thus more detailed information about the dynamics under investigation will be revealed.

{\color{red} The average of inhomogeneous broadening will superpose 2DES with different level spacings $\omega_{a_1a_2}$ between $|a_1\rangle$ and $|a_2\rangle$. When the static disorder is small, the average of inhomogeneous broadening will not essentially modify our main findings, as shown in Fig.~\ref{fig:disorder}(a). As the static disorder increases, the homogeneous broadening will be enlarged due to the average of inhomogeneous broadening, as shown in Fig.~\ref{fig:disorder}(b). Therefore, we may safely arrive at the conclusion that the anomalously-reduced homogeneous broadening can be still observed as long as $\sigma<2.6 \varGamma_1$, which is within the experimental observation $\sigma=1.4 \varGamma_1$ \cite{Novoderezhkin2011}.

\begin{figure}
\includegraphics[width=8.5cm]{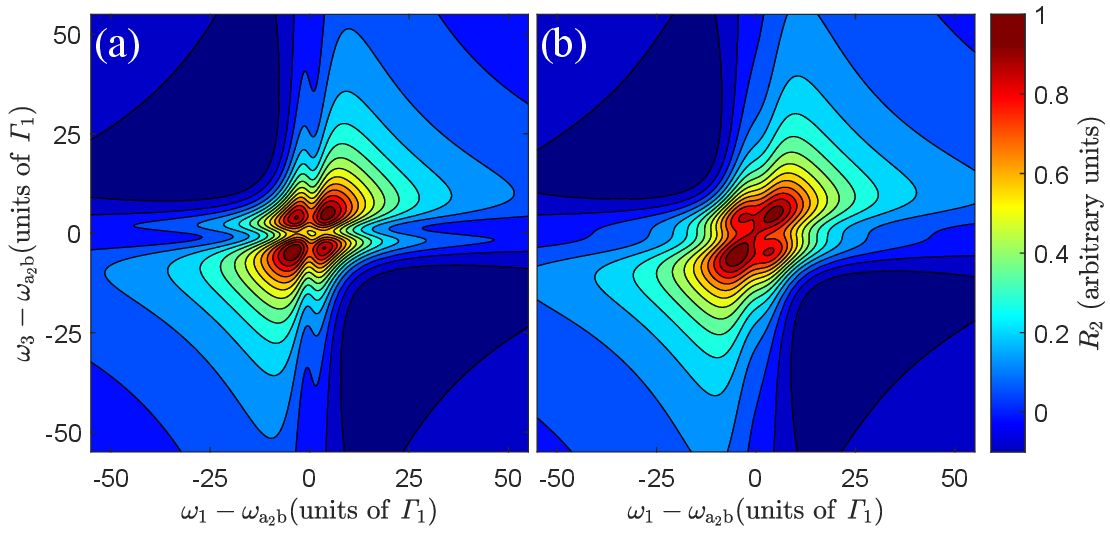}
\caption{2DES with $\omega_{a_{1}b}-\omega_{a_{2}b}=0.18\varGamma_{1}$ when $t_2=3\varGamma_1^{-1}$, $k_BT=5\varGamma_{1}$, $\Omega=9\varGamma_{1}$, and (a) $\sigma = \varGamma_{1}$, (b) $\sigma = 2.6\varGamma_{1}$.}
\label{fig:disorder}
\end{figure}

So far as the average of the interaction between the electric field and the transition dipole moments of molecules is considered, it will superpose 2DES with different widths of the gaps between the peaks in Fig.~\ref{fig:NearResonant}(d). If the electric field is big enough or the interaction between the electric field and the transition dipole moments of molecules are nearly identical, our main findings will not be essentially modified. In order to make the interaction between the electric field and the transition dipole moments of molecules nearly identical, we can initially prepare the sample to be crystal \cite{Zhao2023}.}

\section{Physical Implementation}
\label{sec:Implementation}

To demonstrate our theoretical proposal, we use the energy-level structure of two pheophytins, i.e., PheD1 and PheD2, in the PSII-RC. Their energies are respectively $E_{1}=15030~\mathrm{cm^{-1}}$ and $E_{2}=15020~\mathrm{cm^{-1}}$. The electronic coupling between them is $J=-3~\mathrm{cm^{-1}}$ \cite{ChemistryE2011}. We calculate the new energy levels of the dimer due to the coupling as $\varepsilon_{1}=(E_{1}+E_{2}+\Delta E)/2=15031~\mathrm{cm^{-1}}$ and $\varepsilon_{2}=(E_{1}+E_{2}-\Delta E)/2=15019~\mathrm{cm^{-1}}$ where $\Delta E=\sqrt{(E_{1}-E_{2})^{2}+4J^{2}}$ is the gap between two levels $a_{1}$ and $a_{2}$. In order to realize the four-level configuration in Fig.~\ref{fig:scheme}, we utilize a vibronic mode with frequency $745~\textrm{cm}^{-1}$ with relaxation rate $\varGamma_b=0.1~\mathrm{ps}^{-1}$ \cite{Reimers13SR}, i.e., $|b\rangle=|g\rangle|1\rangle_v$, $|c\rangle=|g\rangle|0\rangle_v$, $|a_1\rangle=|\varepsilon_{1}\rangle|0\rangle_v$, and $|a_2\rangle=|\varepsilon_{2}\rangle|0\rangle_v$, where $|0\rangle_v$ and $|1\rangle_v$ are respectively the ground and first excited state of the vibronic mode. The pure-dephasing rates are respectively $\gamma_{a_{1}c}^{(0)}=\gamma_{a_{1}b}^{(0)}=\gamma_{a_{2}b}^{(0)}=10\varGamma_1$ and $\gamma_{bc}^{(0)}=10\varGamma_b$. The Rabi frequency of the control field is $\Omega=9\varGamma_1$. The downhill population transfer rate is $\varGamma_{1}=10~\mathrm{ps}^{-1}$. Here, the uphill population transfer rate $\varGamma_{2}$ and $\varGamma_{1}$ satisfy the detailed-balance relation, i.e., $\varGamma_{2}/\varGamma_{1}=\textrm{exp}(-\Delta E/k_{B}T)$, where $T$ is the ambient temperature.

\section{Conclusion and Discussion}
\label{sec:Conclusion}
{\color{red}The proposed anomalous effect can be observed for a wide range of parameters, which will be elucidated as follows. First of all, the temperature must be sufficiently high to satisfy $\varGamma_1/\varGamma_2 \approx 1$, which is easily achieved for a molecule with small level spacing, i.e., $\hbar \omega_{a_1a_2}\ll k_BT$. This is because the contributions from the off-diagonal peaks become more significant at higher temperatures. Secondly, the population time $t_2$ must exceed the inverse of the relaxation rate $1/\varGamma_1$, i.e., $t_2\varGamma_1>1$, ensuring that the off-diagonal peaks remain relatively stable over time. The last but not the least, the Rabi frequency of the control field needs to be sufficiently large, as the separation of the sub-peaks introduced by the EIT is proportional to the Rabi frequency. In this study, a significant phenomenon can be observed at $\Omega=9\varGamma_1$.}

In this paper, we theoretically explore the anomalous reduction of homogeneous broadening in 2DES at high temperatures, which is attributed to the detailed balance assisted by the EIT. Compared with lower temperatures, the homogeneous broadening is much narrower at higher temperatures due to the long-lasting off-diagonal peaks, which vanish at the former case.
Since in realistic experiments, $R_2$ can not be separated from the rephasing signal which also contains $R_3$, our discovery still holds when $R_3$ is included. When the static disorder is considered, the homogeneous broadening is remarkably suppressed as long as $\sigma<2.6\varGamma_1$, which is within the experimental observation.

\section*{Acknowledgments}
We thank stimulating discussions with W.-T. He and N.-N. Zhang. This work is supported by Innovation Program for Quantum Science and Technology under Grant No.~2023ZD0300200, Beijing Natural Science Foundation under Grant No.~1202017 and the National Natural Science Foundation of China under Grant Nos.~62105030,~11674033,~11505007, and Beijing Normal University under Grant No.~2022129.

{\color{red}
\appendix
\section{Master Equation for 2-Level System}
\label{sec:appendixA}
	In this appendix, we will derive the master equation for 2-level system \cite{Carmichael1993}. We begin by considering a quantum system $S$ coupled to a reservoir $R$. The Hamiltonian of the total system is
	\begin{equation}
		H=H_{S}+H_{R}+H_{I},
	\end{equation}
	where $H_{S}$ and $H_{R}$ are respectively the Hamiltonian of the system and the reservoir, $H_{I}$ describes the interaction
	between them.
	
	In the interaction picture, the von Neumann equation reads
	\begin{equation}
\frac{d}{dt}\rho_{I}(t)=-\frac{i}{\hbar}\left[H_{I}(t),\rho_{I}(t)\right],\label{eq:2}
	\end{equation}
where $\hbar$ is the reduced Planck constant.
	Formally, the total density matrix $\rho_{I}(t)$ can be given as
	\begin{equation}
		\rho_{I}(t)=\rho_{I}(0)-\frac{i}{\hbar}\int_{0}^{t}dt^{'}\left[H_{I}(t^{'}),\rho_{I}(t^{'})\right].
	\end{equation}
	Inserting it into Eq.~(\ref{eq:2}) and tracing over the degrees of freedom of the reservoir,
	we find
	\begin{equation}
		\frac{d}{dt}\rho_{S}(t)=-\frac{1}{\hbar^{2}}\int_{0}^{t}dt^{'}\textrm{tr}_{R}\left[H_{I}(t),\left[H_{I}(t^{'}),\rho_{I}(t^{'})\right]\right],
		\label{eq:4}\end{equation}
	where we have assumed $\textrm{tr}_{R}\left[H_{I}(t),\rho_{I}(0)\right]=0$.
	
	Assuming that the coupling between the system and the reservoir is
	weak, the backaction of the system on the reservoir can be neglected. Therefore,
	the density matrix of the reservoir $\rho_{R}$ is only negligibly
	affected by the interaction and the state of the total system at time $t$ can be approximated by a tensor product, i.e., the Born approximation \cite{Breuer2002},
	\begin{equation}
		\rho_{I}(t)\approx\rho_{S}(t)\otimes\rho_{R}.\label{eq:5}
	\end{equation}
	By substituting Eq.~(\ref{eq:5}) into Eq.~(\ref{eq:4}), we have
	\begin{equation}
		\frac{d}{dt}\rho_{S}(t)=-\frac{1}{\hbar^{2}}\int_{0}^{t}dt^{'}\textrm{tr}_{R}\left[H_{I}(t),\left[H_{I}(t^{'}),\rho_{S}(t^{'})\otimes\rho_{R}\right]\right].
	\end{equation}
	We perform the Markovian approximation \cite{Breuer2002}, in which the time evolution of the state of the system at time $t$ only depends on the present state $\rho_{S}(t)$, i.e.,
	\begin{equation}
		\frac{d}{dt}\rho_{S}(t)=-\frac{1}{\hbar^{2}}\int_{0}^{t}dt^{'}\textrm{tr}_{R}\left[H_{I}(t),\left[H_{I}(t^{'}),\rho_{S}(t)\otimes\rho_{R}\right]\right].
	\end{equation}
	
	As an example, we consider a two-level system, whose two states are denoted as $|1\rangle$ and $|2\rangle$, with energies $E_{1}$ and $E_{2}$ ($E_{1}<E_{2}$), respectively.
	The Hamiltonian $H_{S}$ can be written as
	\begin{equation}
		H_{S}=\frac{1}{2}\hbar\omega_{A}\sigma_{z},
	\end{equation}
	where $\omega_{A}=(E_{2}-E_{1})/\hbar$, $\sigma_{z}=|2\rangle\langle2|-|1\rangle\langle1|$.
	
	Since the system interacts with a reservoir $R$, the total Hamiltonian is given by \cite{Carmichael1993}
	\begin{equation}
		H_{T}=H_{S}+H_{R}+H_{I},	
	\end{equation}
	where
	\begin{eqnarray}
		H_{R}&=&\sum_{k,\lambda}\hbar\omega_{k}r_{k\lambda}^{\dagger}r_{k\lambda},\\
		H_{I}&=&\sum_{k,\lambda}\hbar(\kappa_{k\lambda}^{\ast}r_{k\lambda}^{\dagger}\sigma_{-}+\kappa_{k\lambda}r_{k\lambda}\sigma_{+}).
	\end{eqnarray}
$\omega_{k}$ is the frequency of $k$th mode of the reservoir with
polarization $\lambda$ and $r_{k\lambda}^{\dagger}$ ($r_{k\lambda}$) being the raising (lowering) operator.
	Here, the coupling between the atom and the $k$th mode of the reservoir with polarization $\lambda$ is
	\begin{equation}
		\kappa_{k\lambda}=-ie^{i\vec{k}\cdot \vec{r}_{A}}\sqrt{\frac{\omega_{k}}{2\hbar\epsilon_{0}V}}\hat{e}_{k\lambda}\cdot \vec{d}_{21}.
	\end{equation}
	The unit polarization vector is $\hat{e}_{\vec{k}\lambda}$. The atom is positioned at $\vec{r}_{A}$
	and $V$ is the quantized volume. $\sigma_{+}=|2\rangle\langle1|=(\sigma_{-})^\dagger$ are the raising and lowering operators of the atom. $\epsilon_{0}$ is the dielectric constant of vacuum. $\vec{d}_{21}$ is the transition dipole moment of the atom.

	For simplicity, we define the following operators
	\begin{eqnarray}
		s_{1}&=&\sigma_{-},\\
		s_{2}&=&\sigma_{+},\\
		\varGamma_{1}&=&\varGamma^{\dagger}=\sum_{k,\lambda}\kappa_{k\lambda}^{\ast}r_{k\lambda}^{\dagger},\\
		\varGamma_{2}&=&\varGamma=\sum_{k,\lambda}\kappa_{k\lambda}r_{k\lambda},
	\end{eqnarray}
	where $s_{i}$'s and $\varGamma_{i}$'s are respectively the operators in the Hilbert space of $S$ and $R$. In
	the interaction picture with respect to $H_{S}+H_{R}$, we have
	\begin{eqnarray}
		\widetilde{s}_{1}(t)&=&\sigma_{-}e^{-i\omega_{A}t},\\
		\widetilde{s}_{2}(t)&=&\sigma_{+}e^{i\omega_{A}t},\\
		\widetilde{\varGamma}_{1}(t)&=&\widetilde{\varGamma}^{\dagger}(t)=\sum_{k,\lambda}\kappa_{k,\lambda}^{\ast}r_{k,\lambda}^{\dagger}e^{i\omega_{k}t},\\
		\widetilde{\varGamma}_{2}(t)&=&\widetilde{\varGamma}(t)=\sum_{k,\lambda}\kappa_{k,\lambda}r_{k,\lambda}e^{-i\omega_{k}t}.
	\end{eqnarray}
	
	Therefore, the interaction Hamiltonian in the interaction picture reads
	\begin{equation}
		\widetilde{H}_{I}(t)=\hbar\sum_{i=1,2}\widetilde{s}_{i}(t)\widetilde{\varGamma}_{i}(t).
	\end{equation}
	$R_{0}$ is the initial density operator of the reservoir. The master equation	is now
	\begin{eqnarray}
		\dot{\widetilde{\rho}} & = & -\sum_{i,j}\int_{0}^{t}dt^{'}\{[\widetilde{s}_{i}(t)\widetilde{s}_{j}(t^{'})\widetilde{\rho}(t^{'})-\widetilde{s}_{j}(t^{'})\widetilde{\rho}(t^{'})\widetilde{s}_{i}(t)]\nonumber\\
		&& \times\langle\widetilde{\varGamma}_{i}(t)\widetilde{\varGamma}_{j}(t^{'})\rangle_{R}+  [\widetilde{\rho}(t^{'})\widetilde{s}_{j}(t^{'})\widetilde{s}_{i}(t)-\widetilde{s}_{i}(t)\widetilde{\rho}(t^{'})\widetilde{s}_{j}(t^{'})]\nonumber\\
&&\times\langle\widetilde{\varGamma}_{j}(t^{'})\widetilde{\varGamma}_{i}(t)\rangle_{R}\},
	\end{eqnarray}
	where we have used the cyclic property of the trace, i.e., $\textrm{tr}(ABC)=\textrm{tr}(CAB)=\textrm{tr}(BCA)$.
	And the two correlation functions are
	\begin{eqnarray}
		\langle\widetilde{\varGamma}_{i}(t)\widetilde{\varGamma}_{j}(t^{'})\rangle_{R} & = & \textrm{tr}_{R}[R_{0}\widetilde{\varGamma}_{i}(t)\widetilde{\varGamma}_{j}(t^{'})],\\
		\langle\widetilde{\varGamma}_{i}(t^{'})\widetilde{\varGamma}_{j}(t)\rangle_{R} & = & \textrm{tr}_{R}[R_{0}\widetilde{\varGamma}_{j}(t^{'})\widetilde{\varGamma}_{i}(t)].
	\end{eqnarray}

	Thus, we have
	\begin{eqnarray}
		\dot{\widetilde{\rho}} & = & -\int_{0}^{t}dt^{'}\{[\sigma_{-}\sigma_{-}\widetilde{\rho}(t^{'})-\sigma_{-}\widetilde{\rho}(t^{'})\sigma_{-}]e^{-i\omega_{A}(t+t^{'})}\langle\widetilde{\varGamma}^{\dagger}(t)\widetilde{\varGamma}^{\dagger}(t^{'})\rangle_{R}\nonumber\\
		&& +  [\sigma_{+}\sigma_{+}\widetilde{\rho}(t^{'})-\sigma_{+}\widetilde{\rho}(t^{'})\sigma_{+}]e^{i\omega_{A}(t+t^{'})}\langle\widetilde{\varGamma}(t)\widetilde{\varGamma}(t^{'})\rangle_{R}\nonumber\\
		&& + [\sigma_{-}\sigma_{+}\widetilde{\rho}(t^{'})-\sigma_{+}\widetilde{\rho}(t^{'})\sigma_{-}]e^{-i\omega_{A}(t-t^{'})}\langle\widetilde{\varGamma}^{\dagger}(t)\widetilde{\varGamma}(t^{'})\rangle_{R}\nonumber\\
		&& + [\sigma_{+}\sigma_{-}\widetilde{\rho}(t^{'})-\sigma_{-}\widetilde{\rho}(t^{'})\sigma_{+}]e^{i\omega_{A}(t-t^{'})}\langle\widetilde{\varGamma}(t)\widetilde{\varGamma}^{\dagger}(t^{'})\rangle_{R}\}\nonumber\\
&&+\textrm{h.c.},
	\end{eqnarray}
	where the correlation functions of the reservoir are explicitly
	\begin{eqnarray}	\langle\widetilde{\varGamma}^{\dagger}(t)\widetilde{\varGamma}^{\dagger}(t^{'})\rangle_{R} & = & 	\langle\widetilde{\varGamma}(t)\widetilde{\varGamma}(t^{'})\rangle_{R} = 0,\\
		\langle\widetilde{\varGamma}^{\dagger}(t)\widetilde{\varGamma}(t^{'})\rangle_{R} & = & \sum_{k,\lambda}|\kappa_{k,\lambda}|^{2}e^{i\omega_{k}(t-t^{'})}\bar{n}(\omega_{k},T),\\
		\langle\widetilde{\varGamma}(t)\widetilde{\varGamma}^{\dagger}(t^{'})\rangle_{R} & = & \sum_{k,\lambda}|\kappa_{k,\lambda}|^{2}e^{-i\omega_{k}(t-t^{'})}[\bar{n}(\omega_{k},T)+1].\nonumber\\
	\end{eqnarray}
	The average photon number of the $k$th mode of the reservoir is
	\begin{equation} \bar{n}(\omega_{k},T)=\textrm{tr}_{R}(R_{0}r_{k,\lambda}^{\dagger}r_{k,\lambda})=\frac{e^{-\hbar\omega_{k}/k_{B}T}}{1-e^{-\hbar\omega_{k}/k_{B}T}},
	\end{equation}
	where $k_{B}$ is the Boltzmann constant, $T$ is the temperature.

	Defining $\tau=t-t^{'}$, we have
	\begin{eqnarray}
		\dot{\widetilde{\rho}} & = & \int_{0}^{t}dt^{'}\{[\sigma_{-}\sigma_{+}\widetilde{\rho}(t-\tau)-\sigma_{+}\widetilde{\rho}(t-\tau)\sigma_{-}]e^{-i\omega_{A}\tau}\langle\widetilde{\varGamma}^{\dagger}(t)\widetilde{\varGamma}(t-\tau)\rangle_{R}\nonumber\\
		&  & +[\sigma_{+}\sigma_{-}\widetilde{\rho}(t-\tau)-\sigma_{-}\widetilde{\rho}(t-\tau)\sigma_{+}]e^{i\omega_{A}\tau}\langle\widetilde{\varGamma}(t)\widetilde{\varGamma}^{\dagger}(t-\tau)\rangle_{R}\}\nonumber\\
&  &+\textrm{h.c.}
	\end{eqnarray}
	
	In the continuum limit, the nonvanishing correlation functions of the reservoir are
	\begin{eqnarray}
		\langle\widetilde{\varGamma}^{\dagger}(t)\widetilde{\varGamma}(t-\tau)\rangle_{R} \!\!& = &\!\! \sum_{\lambda}\int d^{3}kg(k)|\kappa_{k,\lambda}|^{2}e^{i\omega_{k}\tau}\bar{n}(\omega_{k},T),\\
		\langle\widetilde{\varGamma}(t)\widetilde{\varGamma}^{\dagger}(t-\tau)\rangle_{R} \!\!& = &\!\! \sum_{\lambda}\int d^{3}kg(k)|\kappa_{k,\lambda}|^{2}e^{-i\omega_{k}\tau}[\bar{n}(\omega_{k},T)+1],\nonumber\\
	\end{eqnarray}
	where $g(k)$ is the density of states of the reservoir.
	Because
	\begin{equation}
		\lim_{t\rightarrow\infty}\int_{0}^{t}d\tau e^{-i(\omega_{k}-\omega_{A})\tau}=\pi\delta(\omega_{k}-\omega_{A})+iP\left(\frac{1}{\omega_{A}-\omega_{k}}\right),
	\end{equation}
	where $P$ indicates the Cauchy principal value and $\omega_{k}=kc$. Based on the Markovian approximation,
	we replace $\widetilde{\rho}(t-\tau)$ by $\widetilde{\rho}(t)$ and
	obtain
\begin{widetext}
	\begin{eqnarray}
		\dot{\widetilde{\rho}} & = & \sum_{\lambda}\int d^{3}kg(k)|\kappa_{k,\lambda}|^{2}\left[\pi\delta(kc-\omega_{A})+iP\left(\frac{1}{\omega_{A}-\omega_{k}}\right)\right][\bar{n}(kc,T)+1](\sigma_{-}\widetilde{\rho}\sigma_{+}-\sigma_{+}\sigma_{-}\widetilde{\rho})\nonumber \\
		&& +  \sum_{\lambda}\int d^{3}kg(k)|\kappa_{k,\lambda}|^{2}\left[\pi\delta(kc-\omega_{A})+iP\left(\frac{1}{\omega_{A}-\omega_{k}}\right)\right]\bar{n}(kc,T)(\sigma_{+}\widetilde{\rho}\sigma_{-}-\widetilde{\rho}\sigma_{-}\sigma_{+})+\textrm{h.c.}
	\end{eqnarray}
\end{widetext}
	
	To summarize, the master equation for the two-level system reads
	\begin{eqnarray}
		\dot{\widetilde{\rho}} & = & \left[\frac{\gamma}{2}(\bar{n}+1)+i(\varDelta^{'}+\varDelta)\right](\sigma_{-}\widetilde{\rho}\sigma_{+}-\sigma_{+}\sigma_{-}\widetilde{\rho})\nonumber \\
		&& +  \left(\frac{\gamma}{2}\bar{n}+i\varDelta^{'}\right)(\sigma_{+}\widetilde{\rho}\sigma_{-}-\widetilde{\rho}\sigma_{-}\sigma_{+})+\textrm{h.c.},
	\end{eqnarray}
	where the relaxation rate at the zero temperature is
	\begin{eqnarray}
		\gamma & = & 2\pi\sum_{\lambda}\int d^{3}kg(k)|\kappa_{k,\lambda}|^{2}\delta(kc-\omega_{A}),\\
		\Delta & = & \sum_{\lambda}P\left(\int d^{3}k\frac{g(k)|\kappa_{k,\lambda}|^{2}}{\omega_{A}-kc}\right),\\
		\Delta^{'} & = & \sum_{\lambda}P\left(\int d^{3}k\frac{g(k)|\kappa_{k,\lambda}|^{2}}{\omega_{A}-kc}\bar{n}(kc,T)\right)
	\end{eqnarray}
	are the Lamb shifts.
	Therefore, the ratio between the uphill and downhill relaxation rates is determined by the detailed balance as	
	\begin{eqnarray}
		\frac{\varGamma_{2}}{\varGamma_{1}} & = & \frac{\bar{n}}{\bar{n}+1}=e^{-\hbar\omega_{A}/k_{B}T}.
	\end{eqnarray}

\section{Master Equation for 4-Level System}
\label{sec:appendixB}	
	For the four-level system considered in the main text, the Hamiltonian of the system reads
	\begin{equation}
		H_{0} = \sum_{j}\hbar\omega_{j}|j\rangle\langle j|,
	\end{equation}
	where $j=b,a_{1},a_{2},c$, $\omega_{b}<\omega_{c}<\omega_{a_{2}}<\omega_{a_{1}}$. The interaction between the system and the control field is described by the Hamiltonian
	\begin{equation}
		H_\textrm{int} = -\frac{\hbar}{2}\sum_{j=1,2}\Omega_{j}e^{-i\nu_{c}t}|a_j\rangle\langle c|+\textrm{h.c.},
	\end{equation}
	where $\Omega_{j}={\mu_{a_{j}c}\varepsilon_{c}}/{\hbar}$ ($j=1,2$) is the Rabi frequency with $\mu_{a_{i}c}$ being the transition dipole moment between $|a_{i}\rangle$ and $|c\rangle$. $\varepsilon_{c}$ and $\nu_{c}$ are the amplitude and the driving frequency of the control field, respectively.
	In the basis $\{|b\rangle,|a_{1}\rangle,|a_{2}\rangle,|c\rangle\}$, the Hamiltonian of the system can be given in the matrix form as
	\begin{equation}
		\begin{split}
			H & =  H_{0}+H_\textrm{int}\\
			& =  \left(\begin{array}{cccc}
				\hbar\omega_{b} & 0 & 0 & 0\\
				0 & \hbar\omega_{a_{1}} & 0 & -\frac{\hbar}{2}\Omega_{1}e^{-i\nu_{c}t}\\
				0 & 0 & \hbar\omega_{a_{2}} & -\frac{\hbar}{2}\Omega_{2}e^{-i\nu_{c}t}\\
				0 & -\frac{\hbar}{2}\Omega_{1}e^{i\nu_{c}t} & -\frac{\hbar}{2}\Omega_{2}e^{i\nu_{c}t} & \hbar\omega_{c}
			\end{array}\right).
		\end{split}
	\end{equation}
	Furthermore, we assume that the control field is resonant only with the transition between $a_{1}$ and $c$, namely $\Omega_{1}=\Omega$ and  $\Omega_{2}=0$. Then the Hamiltonian is simplified as
	\begin{eqnarray}
		H & = & \left(\begin{array}{cccc}
			\hbar\omega_{b} & 0 & 0 & 0\\
			0 & \hbar\omega_{a_{1}} & 0 & -\frac{\hbar}{2}\Omega e^{-i\nu_{c}t}\\
			0 & 0 & \hbar\omega_{a_{2}} & 0\\
			0 & -\frac{\hbar}{2}\Omega e^{i\nu_{c}t} & 0 & \hbar\omega_{c}
		\end{array}\right).
	\end{eqnarray}
	
	
	As a result, the master equation can be written in a compact form as
\begin{eqnarray}
		\dot{\rho} & = & -i[H,\rho]+\frac{1}{2}\varGamma_{1}(|a_{2}\rangle\langle a_{1}|\rho|a_{1}\rangle\langle a_{2}|-|a_{1}\rangle\langle a_{2}|a_{2}\rangle\langle a_{1}|\rho)\nonumber \\
		&& +  \frac{1}{2}\varGamma_{2}(|a_{1}\rangle\langle a_{2}|\rho|a_{2}\rangle\langle a_{1}|-\rho|a_{2}\rangle\langle a_{1}|a_{1}\rangle\langle a_{2}|)\nonumber \\
&& +\frac{1}{2}\varGamma_{b}(|g\rangle\langle b|\rho|b\rangle\langle g|-|b\rangle\langle g|g\rangle\langle b|\rho)\nonumber \\
        && +  \frac{1}{2}\varGamma_{b}^{\prime}(|b\rangle\langle g|\rho|g\rangle\langle b|-\rho|g\rangle\langle b|b\rangle\langle g|)\nonumber \\
        && +\frac{1}{2}\varGamma_{c}(|g\rangle\langle c|\rho|c\rangle\langle g|-|c\rangle\langle g|g\rangle\langle c|\rho)\nonumber \\
&& +  \frac{1}{2}\varGamma_{c}^{\prime}(|c\rangle\langle g|\rho|g\rangle\langle c|-\rho|g\rangle\langle c|c\rangle\langle g|)+\textrm{h.c.},
	\end{eqnarray}
	where $|g\rangle$ is the ground state.

	Finally, we can obtain the differential equations for the elements of the density matrix as
	\begin{eqnarray}
		\dot{\rho}_{a_{1}a_{1}} \!\!& = &\!\! \frac{i}{2}\Omega\left(e^{-i\nu_{c}t}\rho_{ca_{1}}-e^{i\nu_{c}t}\rho_{a_{1}c}\right)-\varGamma_{1}\rho_{a_{1}a_{1}}+\varGamma_{2}\rho_{a_{2}a_{2}},\\
		\dot{\rho}_{cc} \!\!& = & \!\! \frac{i}{2}\Omega(e^{i\nu_{c}t}\rho_{a_{1}c}-e^{-i\nu_{c}t}\rho_{ca_{1}})-\varGamma_{c}\rho_{cc},\\
		\dot{\rho}_{a_{2}a_{2}} \!\!& = & \!\! -\varGamma_{2}\rho_{a_{2}a_{2}}+\varGamma_{1}\rho_{a_{1}a_{1}},\\
		\dot{\rho}_{a_{1}c} \!\!& = &\!\! -i\omega_{a_{1}c}\rho_{a_{1}c}+\frac{i}{2}\Omega\left(e^{-i\nu_{c}t}\rho_{cc}-e^{-i\nu_{c}t}\rho_{a_{1}a_{1}}\right)-\gamma_{a_{1}c}\rho_{a_{1}c},\nonumber\\
	\end{eqnarray}
where $\omega_{a_{1}c}=\omega_{a_{1}}-\omega_{c}$, $\gamma_{a_{1}c}=\frac{1}{2}\left(\varGamma_{1}+\varGamma_{c}\right)+\gamma^{(0)}_{a_{1}c}$. For the off-diagonal elements of the density matrix, the effect of transverse relaxation rate $\gamma^{(0)}_{a_{1}c}$ should be considered.

In the interaction picture with respect to $H_{0}$,
the master equation reads
	\begin{eqnarray}
			\dot{\tilde{\rho}}_{a_{1}a_{1}} &=& \frac{i}{2}\Omega\left(\tilde{\rho}_{ca_{1}}-\tilde{\rho}_{a_{1}c}\right)-\varGamma_{1}\tilde{\rho}_{a_{1}a_{1}}+\varGamma_{2}\tilde{\rho}_{a_{2}a_{2}},\\
			\dot{\tilde{\rho}}_{a_{1}c} & =& \frac{i}{2}\Omega\left(\tilde{\rho}_{cc}-\tilde{\rho}_{a_{1}a_{1}}\right)-\gamma_{a_{1}c}\tilde{\rho}_{a_{1}c},\\
			\dot{\tilde{\rho}}_{cc} & = & \frac{i}{2}\Omega\left(\tilde{\rho}_{a_{1}c}-\tilde{\rho}_{ca_{1}}\right)-\varGamma_{c}\tilde{\rho}_{cc},\label{eq:rhoCC}\\
			\dot{\tilde{\rho}}_{a_{2}a_{2}} & = &  -\varGamma_{2}\tilde{\rho}_{a_{2}a_{2}}+\varGamma_{1}\tilde{\rho}_{a_{1}a_{1}}.
	\end{eqnarray}
Here, we have used $\nu_{c}=\omega_{ac}$. In other words, we assume that the driving field is resonant with the transition.

Furthermore, we assume that the population relaxation of state $c$ can be neglected. Thus, Eq.~(\ref{eq:rhoCC}) is further simplified as
	\begin{eqnarray}
		\dot{\tilde{\rho}}_{cc} & = & \frac{i}{2}\Omega\left(\tilde{\rho}_{a_{1}c}-\tilde{\rho}_{ca_{1}}\right).
	\end{eqnarray}
	
\section{Rephasing Signal}
\label{sec:appendixC}

	\begin{figure}[htb]
		\centering
		\includegraphics[scale=0.45]{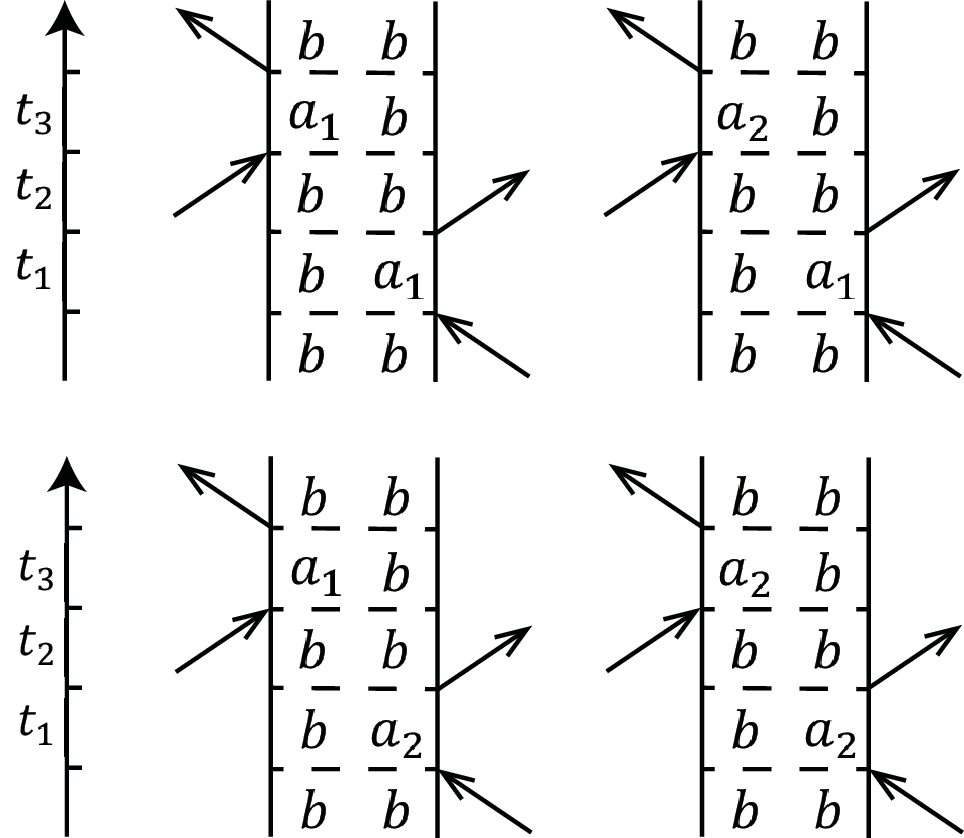}
		\caption{\color{red}Four Feynman diagrams of $R_{3}$.}
		\label{fig:R3}
	\end{figure}

As known to all that in experiments it is a challenge to separate the two pathways, i.e., $R_2$ and $R_3$, in the rephasing direction $k_{s}=-k_{1}+k_{2}+k_{3}$. In this section, we further consider the ground-state bleaching $R_{3}$.
	The two-sided Feynman diagrams of the four possible processes in $R_{3}$ are shown in Fig.~\ref{fig:R3}.
	When considering $R_{3}$, we explore the effect of the population dissipation of energy level $b$ on the whole spectrum. It is found that a similar phenomenon appears in the whole rephasing signal, as shown in Fig.~\ref{fig:rephasing1}.

\begin{figure}[htb]
		\centering
		\includegraphics[width=8.5cm]{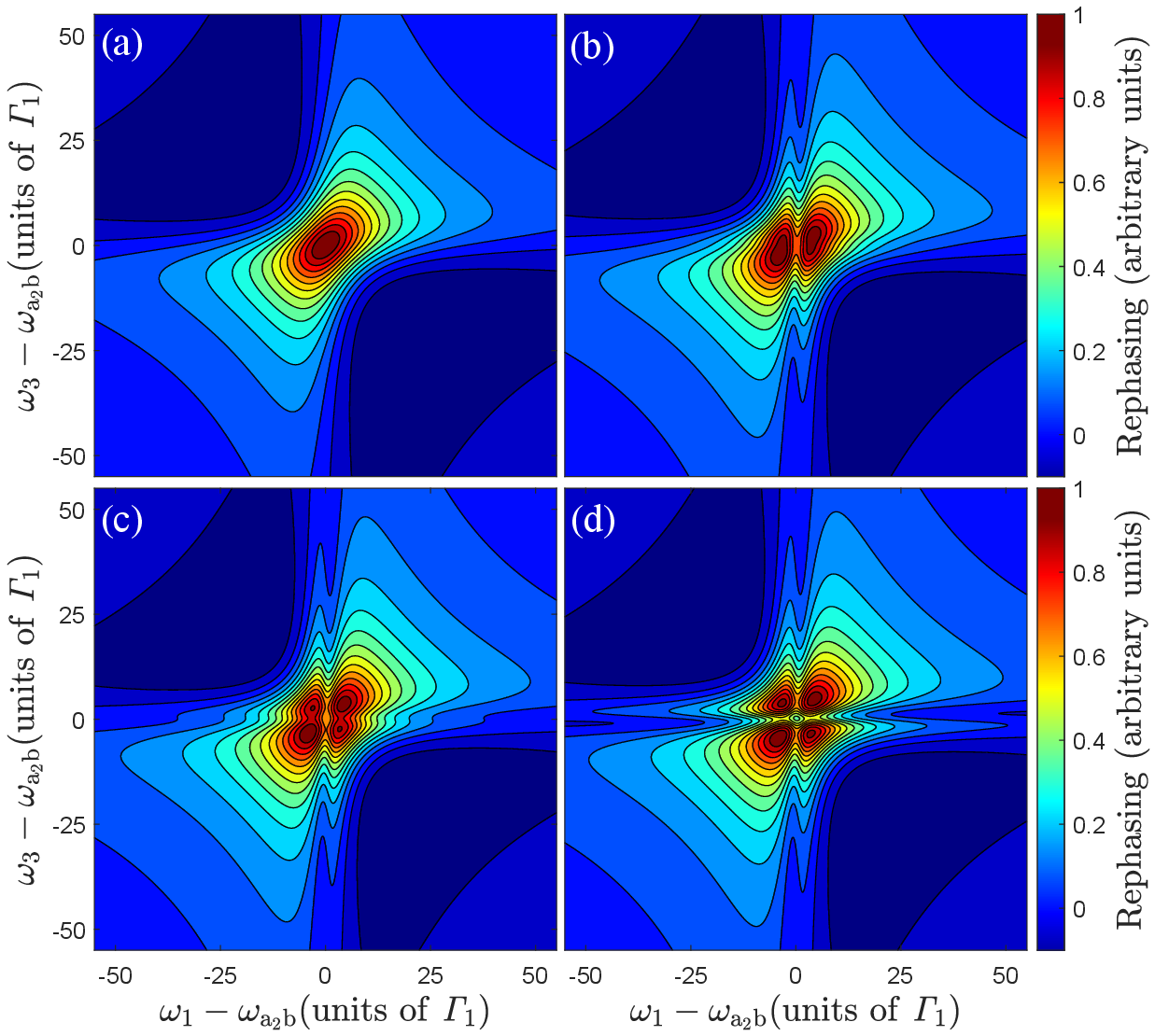}
		\caption{\color{red}Rephasing signal for 2DES when $t_2=3\varGamma_b^{-1}$ and $\omega_{a_{1}b}-\omega_{a_{2}b}=0.18\varGamma_{1}$, and (a) $k_BT=0.01\varGamma_{1}$, $\Omega=0$,
(b) $k_BT=0.01\varGamma_{1}$, $\Omega=9\varGamma_{1}$,
(c) $k_BT=0.1\varGamma_{1}$, $\Omega=9\varGamma_{1}$,
(d) $k_BT=5\varGamma_{1}$, $\Omega=9\varGamma_{1}$.}
		\label{fig:rephasing1}
	\end{figure}

}

\providecommand{\noopsort}[1]{}\providecommand{\singleletter}[1]{#1}%

\end{document}